\documentclass[conference]{IEEEtran}
\IEEEoverridecommandlockouts
\usepackage{cite}
\usepackage{amsmath,amssymb,amsfonts}
\usepackage{algorithmic}
\usepackage{graphicx}
\usepackage{textcomp}
\usepackage{xcolor}

\def\BibTeX{{\rm B\kern-.05em{\sc i\kern-.025em b}\kern-.08em
    T\kern-.1667em\lower.7ex\hbox{E}\kern-.125emX}}
\begin{document}

\title{Misbehavior Detection Using Collective Perception under Privacy Considerations}


\author{\IEEEauthorblockN{Manabu Tsukada\IEEEauthorrefmark{1},
    Shimpei Arii\IEEEauthorrefmark{1}, 
    Hideya Ochiai\IEEEauthorrefmark{1}, and 
    Hiroshi Esaki\IEEEauthorrefmark{1}, 
}
\IEEEauthorblockA{\IEEEauthorrefmark{1}
Graduate School of Information Science and Technology, The University of Tokyo, Tokyo, Japan\\
Email: \textit{\{tsukada, shim, jo2lxq\} @hongo.wide.ad.jp, hiroshi@wide.ad.jp}}
}

\maketitle

\begin{abstract}
  In cooperative ITS, security and privacy protection are essential.
  Cooperative Awareness Message (CAM) is a basic V2V message standard, and misbehavior detection is critical for protection against attacking CAMs from the inside system, in addition to node authentication by Public Key Infrastructure (PKI).
  On the contrary, pseudonym IDs, which have been introduced to protect privacy from tracking, make it challenging to perform misbehavior detection. In this study, we improve the performance of misbehavior detection using observation data of other vehicles. This is referred to as collective perception message (CPM), which is becoming the new standard in European countries.
  We have experimented using realistic traffic scenarios and succeeded in reducing the rate of rejecting valid CAMs (false positive) by approximately 15 percentage points while maintaining the rate of correctly detecting attacks (true positive).
\end{abstract}


\section{Introduction}

Considering the recent development of automated driving technology, the introduction of automated vehicles is being promoted. 
The Society of Automotive Engineers (SAE) in the U.S. has defined six levels of automated driving \cite{SAE_International2021-ay}. 
Cooperative ITS are being developed to achieve a higher level of automated driving because stand-alone autonomous driving has limitations in recognizing the surrounding environment. 

The Cooperative Awareness Message (CAM)\cite{Etsi2014-na} enables the awareness of the surrounding vehicles beyond the limit of the visual range of sensors in a stand-alone automated vehicle by periodically broadcasting information such as position and speed to surrounding vehicles. 
The information in CAM is very important for the safety and efficiency of traffic. If these data are tampered with, or the surrounding environment's awareness is reduced owing to the Sybil attack, it can lead to a severe accident.
Because security is essential in cooperative ITS, Public Key Infrastructure (PKI) is introduced to authenticate nodes and messages.
However, PKI cannot detect and deal with attacks from inside the system. 
For example, the attacker may provoke emergency braking by sending false location information to the space between vehicles.
Therefore, to detect an attack from an authenticated inner node and reduce the damage caused by the attack, Misbehavior Detection in V2X communication has been actively developed \cite{Van_der_Heijden2019-xa}. 
 
In addition, because CAM deals with mobility data, there is a possibility that users' privacy may be violated by tracking vehicles. Considering this, studies have been conducted to ensure the anonymity of cooperative ITS participants by using a short-term identifier known as Pseudonym ID as the source identifier. This is being standardized by ETSI~\cite{Etsi2018-kb}, and being demonstrated in US~\cite{Brecht2018-qp}. 
Pseudonym ID is a crucial element not only in dedicated short-range communications (DSRC) but also in the 5G Automotive Association's (5GAA) report~\cite{5GAA_Automotive_Association2020-zi} on cellular V2X privacy.

In addition to CAM, ETSI is developing Collective Perception Message (CPM)~\cite{Etsi_undated-tu}, a message standard containing information on objects detected by sensors installed in vehicles and roadside units. The CPM enhances the perception of the surrounding environment by compensating for communication blind spots using sensors. 

This study aims at enhancing V2X security by detecting tampered CAM under the privacy-preserving environment, assuming pseudonym IDs are deployed. 
We propose a misbehavior detection method by leveraging the CPM information into the CAM data validation process. 
Our proposed method improves the detection ratio of malicious CAM while reducing the faulty detection ratio. 
The rest of the study is organized as follows. 
Section~\ref{sec:related} presents an overview of the related studies. 
Section~\ref{sec:issues} describes the issues and objectives of the study. 
Section~\ref{sec:proposed} provides our proposal. 
Subsequently, section~\ref{sec:simulation} describes our implementation of simulation and the simulation setting. 
Section~\ref{sec:evaluation} shows the evaluation results of the simulation. 
Finally, section~\ref{sec:conslusion} concludes the study and provides recommendations for future studies. 

\section{Related works} \label{sec:related} 
In VANETs, there are many threats \cite{Raya2005-gd}\cite{Parno2005-pq}\cite{Lu2019-ow} that not only reduce efficiency, but these threats also affect security. Such threats include Sybil attacks, which create fake entities, and attacks that alter parts of the outgoing message. 
%
In the literature, researchers proposed wireless signals-based geolocation verification methods. 
In Xiao et al.\cite{Xiao2006-ap}, Sybil attacks are detected by analyzing the distribution of wireless signals and verifying whether the location information in the message is correct.
Golle et al.~\cite{Golle2004-tx} proposed an algorithm using the strength of the wireless signal to determine the location of a target without depending on whether it has information on effective isotropic radiated power (EIRP). 
Since these schemes do not depend on the node identifier, it works under Pseudonym environment. 

Other works employ recursive Bayesian estimation for misbehavior detection. 
In the study by Sun et al. \cite{Sun2017-td}, the position of the message sender is estimated using a combination of the angle of arrival and Doppler velocity measurements of the received signal as inputs to the extended Kalman filter. 
Bi${\beta}$meyer et al.\cite{Bismeyer2012-bo} use a particle filter to estimate the state of the message sender under pseudonym ID environment. 
Jaeger et al. \cite{Jaeger2012-zv} also considers the pseudonym and determines whether there is a plausible Kalman filter state among several Kalman filter states of the vehicle to be detected. 

Recently, research has emerged that uses machine learning (ML) to perform misbehavior detection.
In \cite{Singh2019-uk}, the authors used ML techniques on the Vehicular Reference Misbehavior dataset (VeReMi)\cite{Van_der_Heijden2018-yn} to detect misbehavior with high accuracy.
\cite{Uprety2021-se} proposed a privacy-preserving misbehavior detection system for the Internet of Vehicles using Federated ML.

Some research has been done on misbehavior detection for attacks against CPM.
In~\cite{Ambrosin2019-sk}, the authors provide a generic design framework that is independent of perception algorithms and analyzes the misbehavior detection system based on the framework against ghost vehicle attacks in the mixed environment. 
MISO-V (Multiple Independent Sources of Observations over V2X)~\cite{Liu2021-lt} addresses the data reliability challenges of CPM by exploiting the inherently overlapping nature of perceptual observations from multiple vehicles to verify the semantic correctness of V2X data.

In the study by Azuma et al. \cite{azuma2017, azuma2018}, we proposed a mechanism for detecting misbehavior by mutually monitoring the vehicle positions when the vehicles upload their own vehicle data to the cloud.

In the above, we overviewed wireless signal-based, recursive Bayesian estimation-based, and ML-based misbehavior detection methods for CAM. Most of them rely on persistent IDs, and there are few targeting pseudonym ID environments. We also investigated misbehavior detection for CPMs, but none of them used CPMs for CAM verification under a privacy-preserving scenario.

\section{Issues and Objectives}\label{sec:issues}

In this study, we assume pseudonym IDs and improve the performance of misbehavior detection through data verification of CAM using CPM. The following sub-sections are the description of the two issues we will address in this study.

\subsection{Adaptation to pseudonym IDs}
Many studies have examined consistency and node reliability verifications based on the assumption that vehicle IDs do not change.
In this study, we assume pseudonym IDs for data verification of CAMs for protecting privacy.
Considering this assumption, it is impossible to identify a particular vehicle by making individual nodes resolve pseudonyms globally. Therefore, it is necessary to have a misbehavior detection method that resolves the pseudonym ID locally or is not affected by the pseudonym ID itself. 

\subsection{Performance improvement of misbehavior detection}
The performance improvement of misbehavior detection in this study aims to reduce the probability of misdetecting a valid message as an attack message (false positive) in CAM data verification while maintaining a successful detection of an attack as an attack (the true-positive rate). 
Particularly, we aim to improve the false-positive rate while maintaining the true-positive rate compared to the conventional methods that use only CAM.
In general, it is possible to increase the true-positive rate by increasing the false-positive rate; however, an increase in the false-positive rate decreases the number of valid CAMs and reduces the effectiveness of messages.
Therefore, it is essential to reduce the false-positive rate while maintaining the true-positive rate. 

\section{Proposed method}\label{sec:proposed}
\begin{figure*}[tb]
  \begin{center}
    \begin{tabular}{c}
      \begin{minipage}{0.5\hsize}
          \centerline{\includegraphics[width=\linewidth,clip]{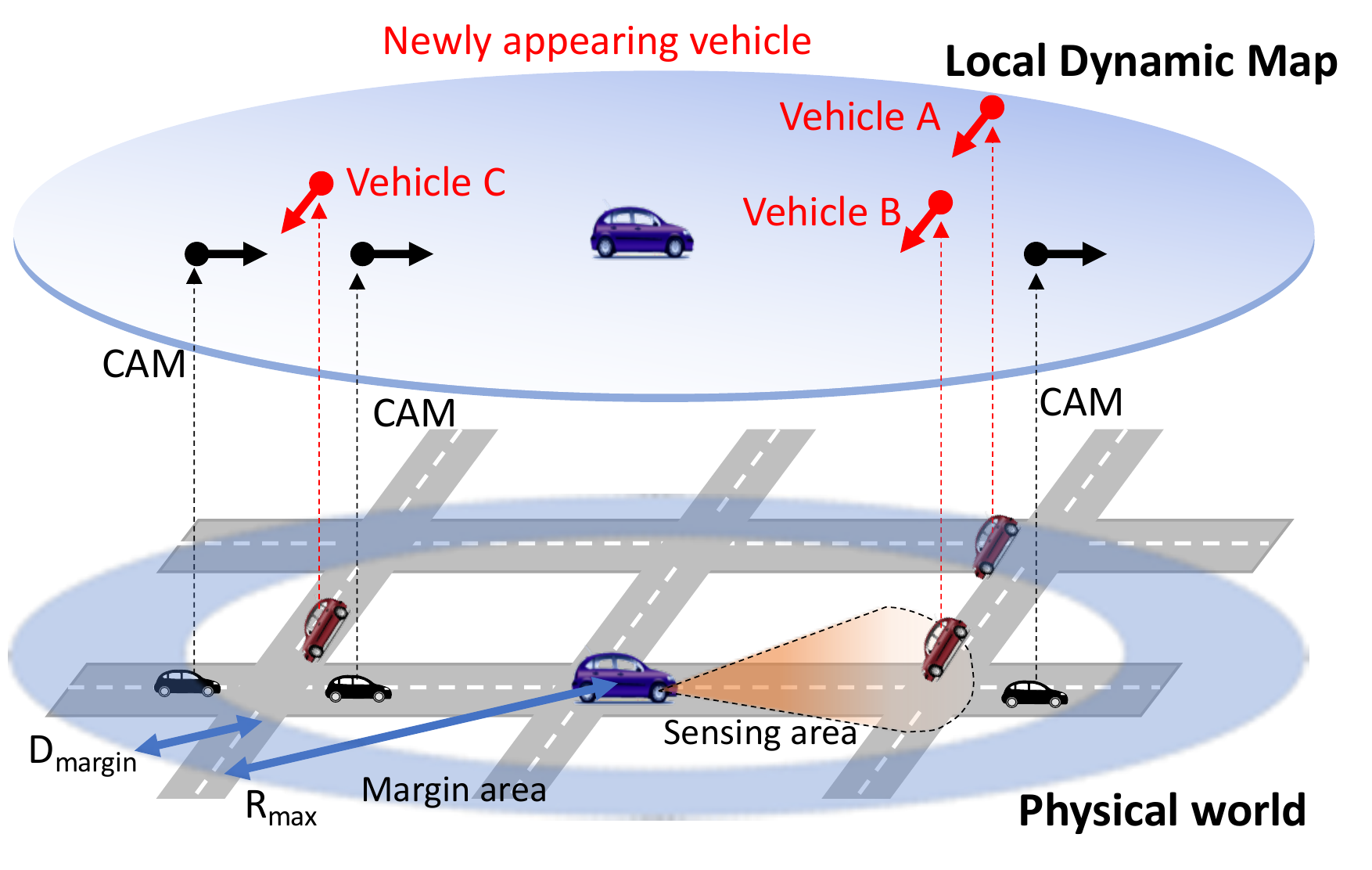}}
          \caption{CAM verification under privacy considerations}
          \label{fig:CAM_verification}
      \end{minipage}
      \begin{minipage}{0.5\hsize}
          \centerline{\includegraphics[width=\linewidth,clip]{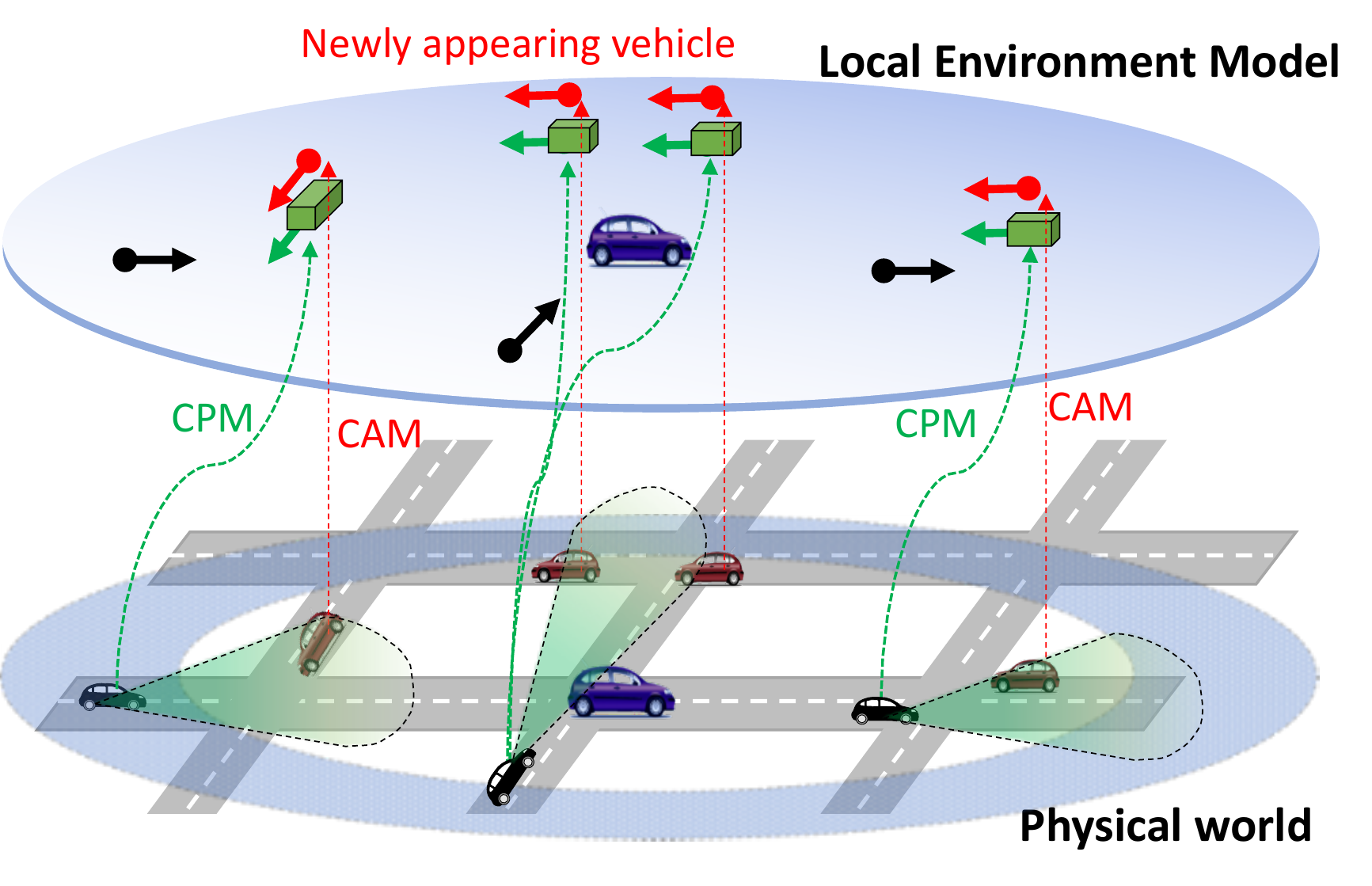}}
          \caption{CAM verification using CPM data}
          \label{fig:CPM_verification}
      \end{minipage}
    \end{tabular}
  \end{center}
\end{figure*}

\subsection{Overview}

Considering the ITS Station Architecture \cite{ISO-21217-CALM-Arch}, the surrounding vehicles, learned by receiving CAMs, are added to the local dynamic map (LDM)~\cite{ETSI-EN-302-895-LDM}. 
Figure~\ref{fig:CAM_verification} shows the CAM verification process. The lower and upper parts represent the physical world and the LDM of the central blue vehicle, respectively. 
The blue car knows the real-time positions of other vehicles in the wireless single-hop by receiving CAMs. 
The black arrows on the LDM show the accepted CAMs, and the red arrows show newly appearing vehicles that are targets of the verification. 
When adding these received CAMs to the LDM, we perform misbehavior detection to accept or reject the CAM.
The Kalman filter is an efficient and well-known solution to estimate time-varying quantities from discrete, error-prone observations such as tracking objects. 
Similar to previous studies~\cite{Stubing2010-hq,Jaeger2012-zv}, we use the Kalman filter to estimate the future motion of the vehicle, and we have used it as the basis for the mobility verification. 
The vehicle has a list of Kalman filters of the surrounding vehicles, and it uses the list to evaluate the received CAM. 

When a vehicle receives CAMs, the valid cases are categorized into three:
1) reception from a station ID that is already known, 
2) reception from a changed pseudonym ID of the known vehicle, and 
3) reception from a new vehicle.

First, when the vehicle receives CAM from a known vehicle ID, its CAM is plausible if it does not deviate from the Kalman filter associated with the sender's vehicle ID. Therefore, it can be accepted. 
Second, when the pseudonym ID is changed, the receiving vehicle holds the Kalman filter associated with the vehicle ID before the change. 
Therefore, if all the Kalman filters in the list are compared to the received CAM, the output of the Kalman filter corresponding to the vehicle ID before the pseudonym change will be the closest value. 
If this value is plausible, it can be detected as a change in the pseudonym ID.

Finally, it is necessary to consider the case where the two cases above are not applied, receiving CAM from a new vehicle. 
Because new vehicles often appear from outside the radio communication range, we define the margin area, which is calculated by the maximum communication range $R_{max}$ and margin distance $D_{margin}$. Moreover, we consider the CAMs transmitted from this area as new valid vehicles (Vehicle A in Fig.~\ref{fig:CAM_verification}). 
Considering some cases, new CAMs from nearby vehicles are received owing to the shielding of buildings or heavy vehicles. 
In such a case, if a matching vehicle can be observed in the vehicle's sensor, it is treated as a new valid vehicle (Vehicle B in Fig.~\ref{fig:CAM_verification}). 
Unfortunately, new vehicles appearing outside the margin area and sensing area cannot be considered as new valid vehicles, as demonstrated by Vehicle C in Fig.~\ref{fig:CAM_verification}. 
Therefore, regardless of whether they are valid or attack messages, all such cases will be rejected. 

To solve this problem, as shown in Fig.~\ref{fig:CPM_verification}, we utilize the Local Environment Model, which holds sensor information of the surrounding area, and the LDM. We employ the sensing capabilities of the surrounding vehicles using the CPM to expand the area of new vehicles that can be verified.
The proposed method uses the sensor-perceived position and velocity information of the object contained in the CPM for generating a new Kalman filter to track a new vehicle more efficiently. 

However, the perceived object of the CPM is used for Kalman filter generation only when the CPM is trustworthy. 
Therefore, the source information of the CPM should have a verification process as well as the CAM verification. Furthermore, we set an upper bound on the sensor's perception range and verify whether the distance between the location of the perceived object and the location of the source does not exceed this upper bound. 
Consequently, even if an attacker tampers with the location information about the perceived object of the CPM, we can limit the range of influence. 

\begin{figure*}[tb]
    \centerline{\includegraphics[width=0.7\linewidth,clip]{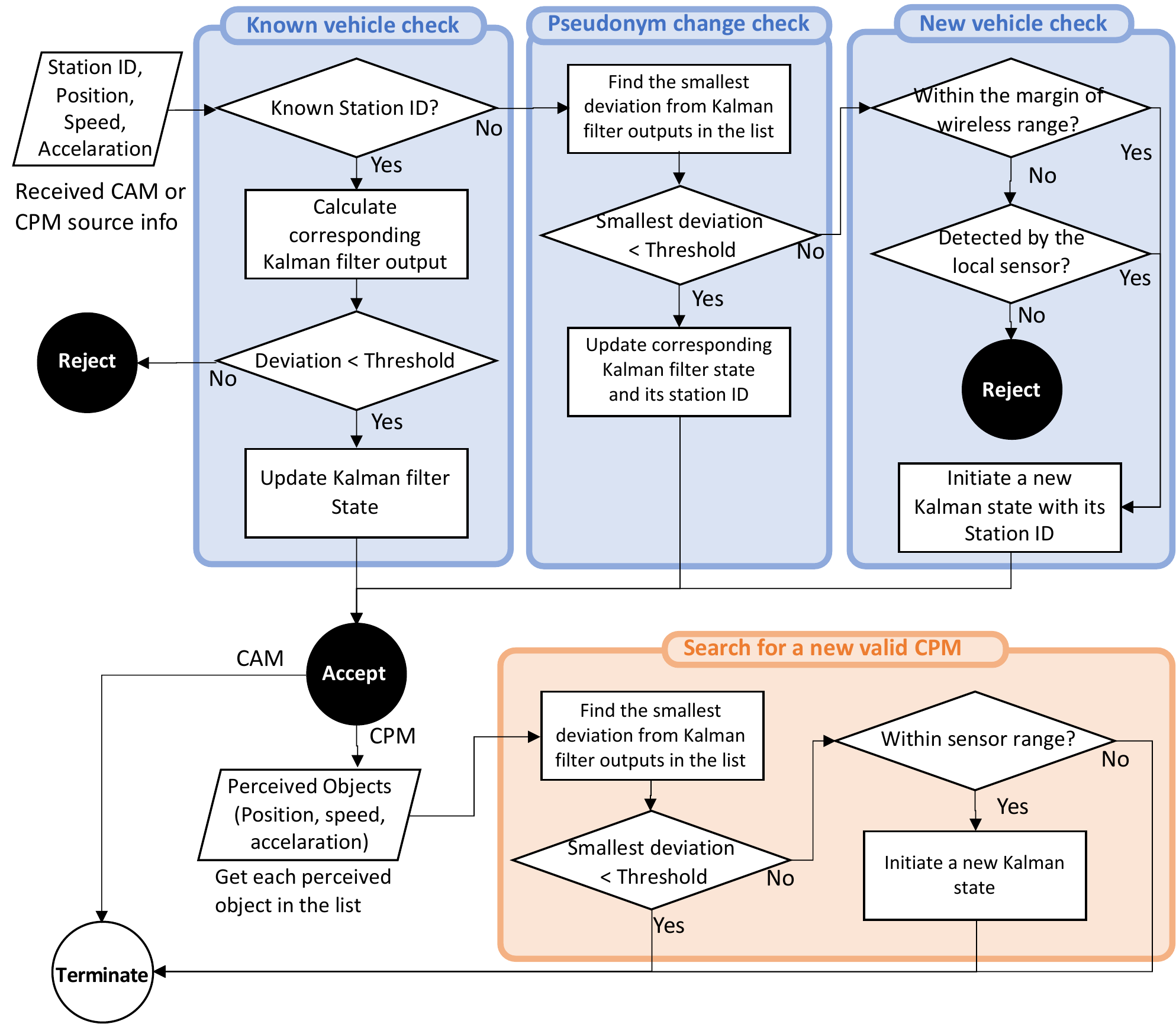}}
    \caption{Flowchart of the proposed method}
    \label{fig:flow1}
  \end{figure*}

Figure \ref{fig:flow1} shows the flowchart of the proposed method. In the following subsections, we describe the processing of CAM and CPM.
The upper blue flow is applied to the CAM and CPM, whereas the lower orange part is applied only to CPM. 

\subsection{CAM processing}\label{subsec:cam}
If the received message is a CAM, our system extracts the source station ID from the ITS Packet Data Unit (PDU) header and extracts the position, speed, and acceleration information from the basic and High Frequency (HF) containers. 
First, it checks whether this CAM is from a known vehicle.
If the Kalman filter corresponding to the retrieved ID exists, 
it uses the acceleration as the input of the Kalman filter for verification. 
Thereafter, we accept the CAM if the position and velocity of CAM do not deviate from the output of the Kalman filter. 
Subsequently, if the Kalman filter corresponding to the retrieved ID does not exist, it checks whether the pseudonym ID has changed. 
Consequently, it verifies the position and velocity retrieved from the CAM using the outputs of all the Kalman filters in the list and selects the one with the smallest difference.
If the value is within the threshold, it updates the state and ID of the corresponding Kalman filter and accepts the CAM. 

If none of the cases above holds, we check if the vehicle newly appears. 
First, it evaluates whether the location information is within the margin area or whether a sensor detects it. If any of the above is true, it accepts the CAM and generates a new Kalman filter; otherwise, the CAM is rejected.

Considering all the processes, we use the Kalman filter used for verification within the time difference threshold between the received time of the message and the last updated time of the Kalman filter to prevent loss of accuracy and incorrect update of IDs.

\subsection{CPM processing}
If the received message is a CPM, our system first extracts the source station ID from the ITS PDU header in the CPM and the sender's location, speed and extracts acceleration information from the management and station data containers. 
Using the retrieved information, we perform the same verification as CAM in Section~\ref{subsec:cam}. 
If the sender information of the CPM is accepted, the subsequent step is to verify the information of the object perceived by the source vehicle (the lower part of Fig.~\ref{fig:flow1}).

When there are detected objects in the CPM, it retrieves the information about all the perceived objects (position information, velocity, and acceleration) from the perceived object container. 
The location, direction of movement, velocity, and acceleration of the perceived object are represented in the local coordinate system, where the sender is the reference.
Therefore, it converts the sensor data into a geodetic reference system (i.e., latitude and longitude) using the sender information retrieved from the management and station data containers. 

Thereafter, it evaluates the CPM using all the Kalman filters in the list. 
This evaluation process is similar to the pseudonym change check described in Section~\ref{subsec:cam}, where each output of all the Kalman filters is used to verify the location and velocity of the perceived object. 
If the smallest deviation is smaller than the threshold, the corresponding Kalman filter exists.
However, the information on the perceived object is indirect; therefore, we do not update the corresponding Kalman filter because it may disarrange the temporal order of the direct information such as CAM or the sender information of the CPM. 
Supposing there is no Kalman filter corresponding to the perceived object information, a new Kalman filter is generated as a new vehicle after verifying whether it is in the target area of the CPM sender vehicle's sensor. 

\section{Simulation Experiment}\label{sec:simulation}
\subsection{System model}
\begin{figure}[tb]
    \centerline{\includegraphics[width=\linewidth,clip]{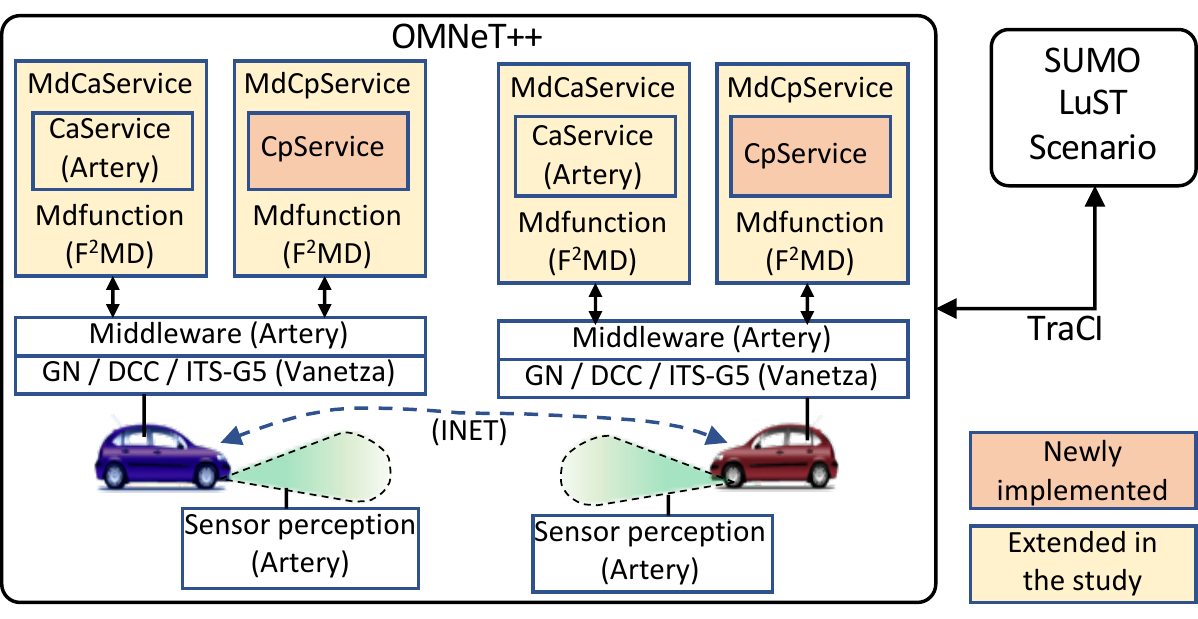}}
    \caption{Simulation architecture}
    \label{fig:archtecture}
\end{figure}

The architecture of the simulation is shown in Figure \ref{fig:archtecture}.

In this study, we used SUMO~\footnote{https://www.eclipse.org/sumo/} as an urban traffic simulator and OMNeT++\footnote{https://omnetpp.org/} as a network simulator. 
In addition, a V2X communication framework running on OMNeT++ which integrates Veins\footnote{https://veins.car2x.org/}, a framework for vehicle communication, INET\footnote{https://inet.omnetpp.org/}, a framework for wired, wireless, and mobile communication, and Vanetza\footnote{https://www.vanetza.org/}, an ITS protocol implementation, were used for implementation. 
Artery\footnote{http://artery.v2x-research.eu/}, a V2X communication framework running on OMNeT++ which integrates Vanetza, an ITS protocol implementation, and F${}^\text{2}$MD~\cite{f2md-journal} provides various functions to simulate the misbehavior detection mechanism. 
Regarding the implementation of the Kalman filter, we used the module provided by F${}^\text{2}$MD. In addition, considering the sensor functions in the vehicle, we used the module provided by the Artery.

Considering each vehicle node, we used the \textit{CaService} module already provided by the Artery and newly implemented the \textit{CpService} module according to the flow in Annex D of \cite{Etsi_undated-tu}. However, the segmentation mechanism is not implemented for \textit{CpService}, and the maximum number of objects included in the perceived object container is set to five.
In addition, to operate the attack generation and misbehavior detection during CAM and CPM transmission and reception, we extended \textit{CaService} and \textit{CpService} to \textit{MdCaService} and \textit{MdCpService}, respectively, using the functions provided by F${}^\text{2}$MD.

\subsection{Parameters and Traffic model}
  

Table \ref{table:param} shows the main simulation parameters. 
All vehicle nodes can send and receive both CAM and CPM messages. Two sensors mounted on the vehicle were evaluated: a sensor with 60$^\circ$ field of view with a range of 80 m in front, and a 360$^\circ$ sensor with a radius of 80 m.

\begin{table}[tb]
	\begin{center}
		\caption{Simulation parameters}
		\begin{tabular}{|c|c|} \hline
			Parameter & Configuration \\ \hline \hline
			Network Simulation Time & 1000 sec \\
			Traffic Simulation Time & 1000 sec  \\ \hline
			Pseudonym Change Period & 100 sec \\ \hline
			Bitrate & 6 Mbps \\
			Tx Power & 200 mW \\
			Carrier Frequency & 5.9 GHz \\ \hline
			CAM DCC Profile & DP2 \\
			CAM Channel & CCH \\
			CPM DCC Profile & DP3 \\
			CPM Channel & CCH \\ \hline
		\end{tabular}
		\label{table:param}
	\end{center}
\end{table}

To simulate the traffic scenario in an urban traffic environment, we used the Luxembourg SUMO Traffic (LuST) Scenario~\cite{codeca2017luxembourg}
, which reflects realistic mobility patterns in Luxembourg, and its reduced version, the LuST Mini Scenario. 
The LuST Scenario is a traffic scenario characterized by a large amount of traffic on the expressway. In contrast, the LuST Mini Scenario is a traffic scenario with only normal roads, not including the expressway.
The maximum number of vehicles that appear in the simulation time is 243 and 105 in the LuST and LuST Mini Scenarios. In all the scenarios, the data were recorded every 50 s.


\subsection{Attacker model}
This study assumes that an attacker falsifies the location information, latitude, and longitude contained in the basic container in the CAM. 
Considering our simulation, a node becomes an attacker with a probability of 10\% of all vehicle nodes loaded in SUMO.
The attacker randomly adds or subtracts 30\% of the CAMs it sends, corresponding to a distance of approximately 3 m to 40 m (considering the LuST Mini Scenario).
Because the width of a typical roadway is 3.5 m, we set the minimum deviation to approximately 3 m. 
The stopping distance of a car traveling at 60 km/h is approximately 37 m, and we assume that the distance between cars traveling at approximately 60 km/h is approximately 40 m. The attacker intends to provoke emergency braking by sending false location information to the space between vehicles. 

The above attack is realized by sending the latitude and longitude information inserted in the basic container by adding random values to the original values. 
The random values range between from 0.00003$^\circ$ and 0.00050$^\circ$ for the LuST scenario and between 0.00003$^\circ$ and 0.00030$^\circ$ for the LuST Mini scenario.
Considering the LuST scenario, the average distance between vehicles is expected to be larger because the speed on the highway is faster than that of the traffic on other roads. Therefore, we set the range of latitude and longitude to be changed larger than that of the Lust Mini scenario by 0.00020$^\circ$.

\section{Evaluation} \label{sec:evaluation}

\subsection{LuST Scenario}


\begin{figure*}[t]   
  \begin{center}
    \begin{tabular}{c}
      \begin{minipage}{0.25\hsize}
        \begin{center}
          \centerline{\includegraphics[width=\linewidth,clip]{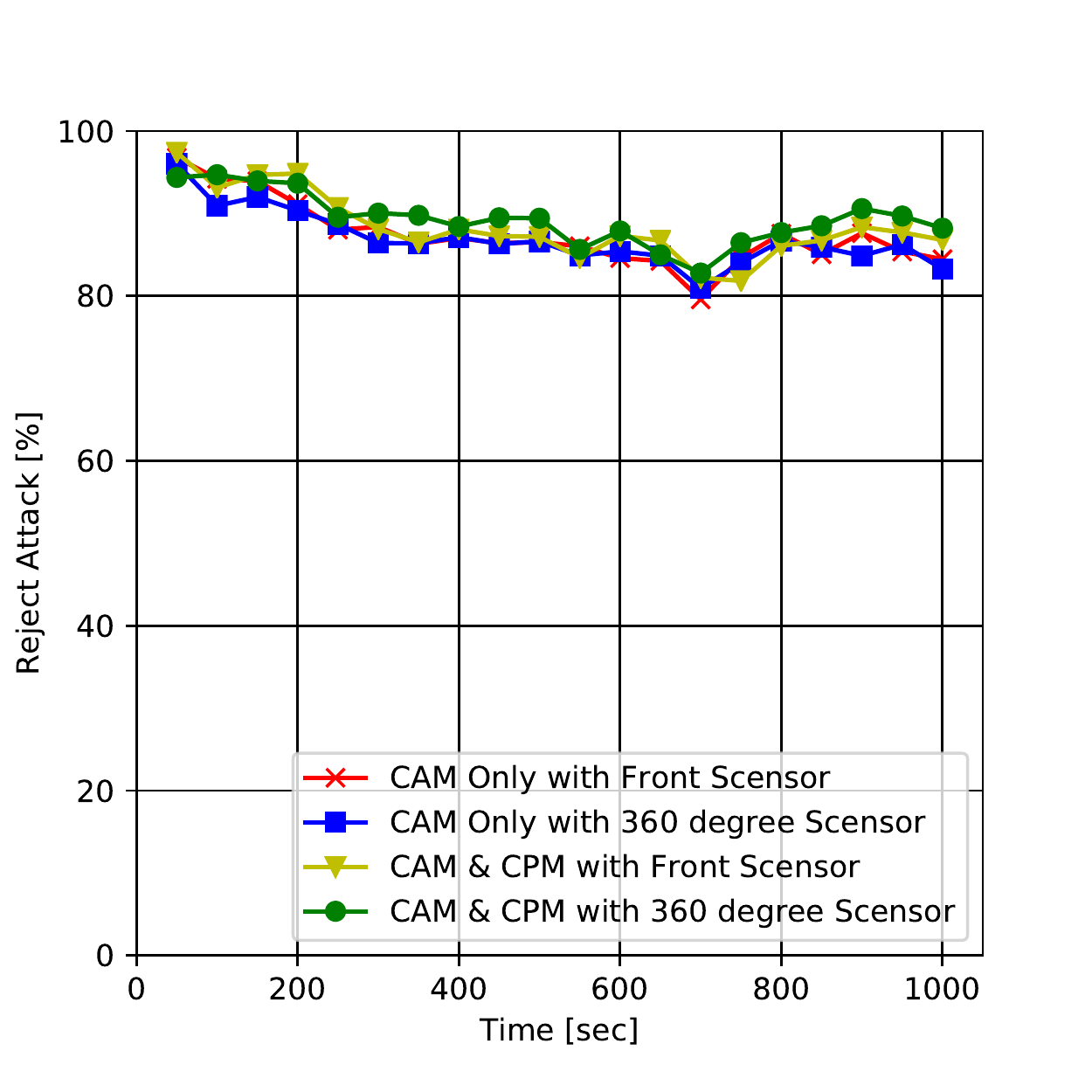}}
          \caption{True positive in LuST}
          \label{fig:lust_high_TP}
          \end{center}
      \end{minipage}
      \begin{minipage}{0.25\hsize}
        \begin{center}
          \centerline{\includegraphics[width=\linewidth,clip]{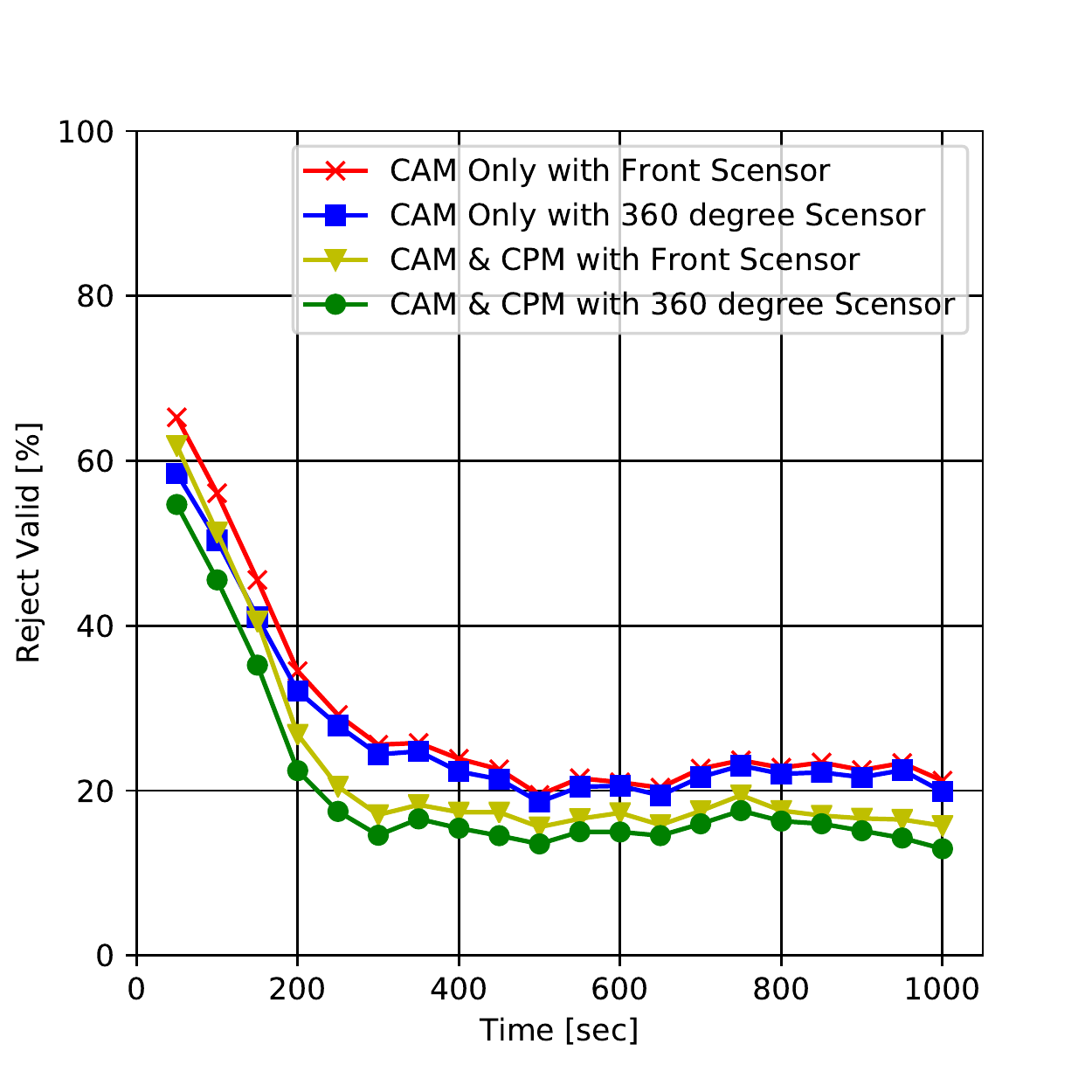}}
          \caption{False positive in LuST }
          \label{fig:lust_high_FP}
          \end{center}
      \end{minipage}
      \begin{minipage}{0.25\hsize}
        \begin{center}
          \centerline{\includegraphics[width=\linewidth,clip]{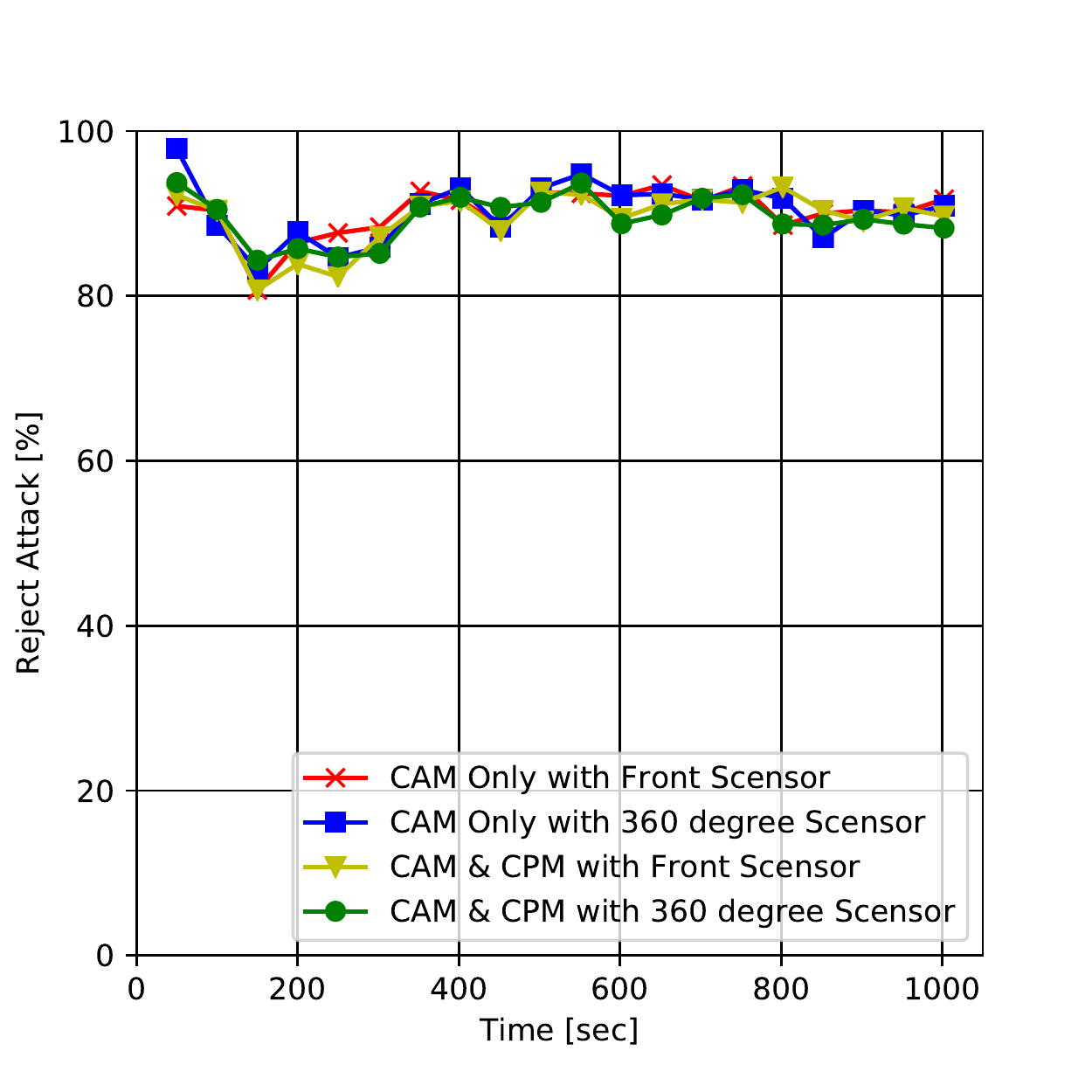}}
          \caption{True positive in LuST Mini}
          \label{fig:mini_high_TP}
        \end{center}
      \end{minipage}
      \begin{minipage}{0.25\hsize}
        \begin{center}
          \centerline{\includegraphics[width=\linewidth,clip]{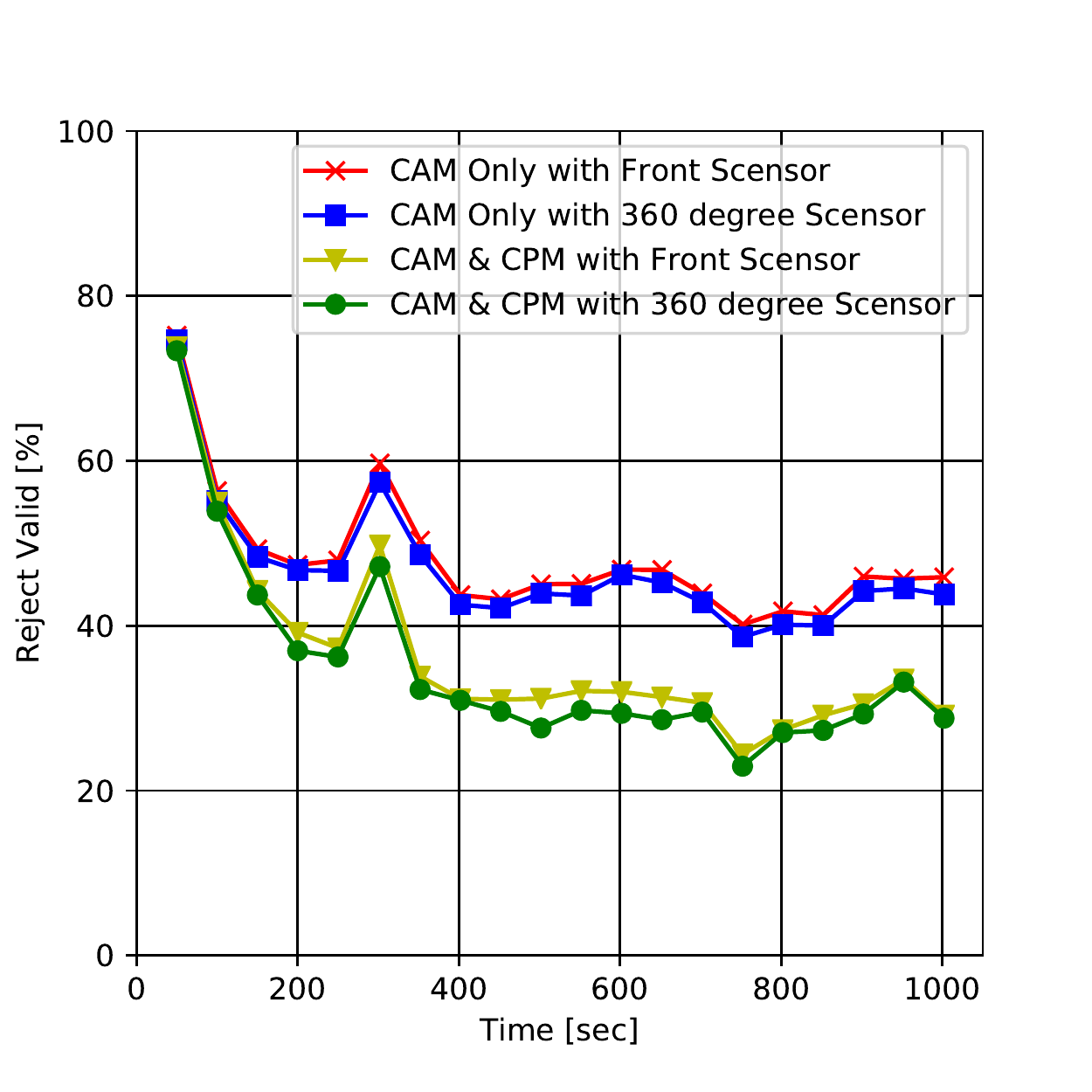}}
          \caption{False positive in LuST Mini}
          \label{fig:mini_high_FP}
        \end{center}
      \end{minipage}
    \end{tabular}
    \end{center}
\end{figure*}

Fig.~\ref{fig:lust_high_TP} and \ref{fig:lust_high_FP} show the simulation results for the true and false positive rates in the LuST scenario. 
Moreover, Table \ref{table:lust_high} is a summary of the average and worst values shown in Fig.~\ref{fig:lust_high_TP} and the average and best values shown in Fig.~\ref{fig:lust_high_FP}.

At the beginning of the experiment, Fig.~\ref{fig:lust_high_TP} and ~\ref{fig:lust_high_FP} show that all the CAMs are rejected, excluding the messages transmitted from the wireless range margin or sensor range because there is no list of Kalman filters. 
Therefore, both valid CAMs and attack CAMs are rejected with high probability. 
Fig.~\ref{fig:lust_high_TP} and ~\ref{fig:lust_high_FP} show that the system becomes a steady state at 500 s after the experiment. 

Fig.~\ref{fig:lust_high_FP} shows that the false-positive rate improved by approximately 6 percentage points on the average for both the front and 360$^\circ$ sensors when using the proposed method. This result also shows that utilizing CPM, which is the perception of other vehicles, is more effective than upgrading the vehicle's sensors from front 60$^\circ$ sensors to 360$^\circ$ sensors. 
Furthermore, whereas the proposed method improves the false-positive rate, it also keeps the true positive rate at the same level as the CAM-only method, as shown in Fig.~\ref{fig:lust_high_FP}.

\begin{table}[tb]
	\begin{center}
		\caption{True- and false-positive rates in the LuST Scenario}
		\begin{tabular}{|c|r|r|r|r|} \hline
			& \multicolumn{2}{c|}{Reject Attack [\%]} & \multicolumn{2}{c|}{Reject Valid [\%]}\\
			\cline{2-5}
            & \multicolumn{1}{c|}{Average} & \multicolumn{1}{c|}{Worst} & \multicolumn{1}{c|}{Average} & \multicolumn{1}{c|}{Best} \\ \hline 
            CAM/Front & 86.0 & 79.6 & 24.0 & 19.5 \\ \hline 
            CAM/360 degree & 85.8 & 80.9 & 22.8 & 18.6 \\ \hline 
            CAM\&CPM/Front & 87.0 & 81.8 & 18.2 & 15.6 \\ \hline 
            CAM\&CPM/360 degree & 88.3 & 82.8 & 16.1 & 12.9 \\ \hline 
		\end{tabular}
		\label{table:lust_high}
	\end{center}
\end{table}

\subsection{LuST Mini Scenario}
Fig.~\ref{fig:mini_high_TP} and \ref{fig:mini_high_FP} show the simulation results for the true- and false-positive rates in the LuST Mini scenario. 
Moreover, Table \ref{table:mini_high} is a summary of the average and worst values shown in Fig.~\ref{fig:mini_high_TP} and the average and best values shown in Fig.~\ref{fig:mini_high_FP}.
Similar to the LuST scenario, the LuST Mini scenario reached a steady-state approximately 500 s after the experiment.

Fig.~\ref{fig:mini_high_FP} shows that the proposed method improves the false-positive rates by 14.0 percentage points on the average, regardless of the type of sensor (from 45.0\% to 31.0\% for the front sensor, and from 43.6\% to 29.6\% for the 360º sensor, according to Table~\ref{table:mini_high}).
Furthermore, the best case shows 15.6 and 15.7 percentage points improvement, from 40.1\% to 24.4\% and from 38.6\% to 23.0\%, considering the front and 360º sensors, respectively.
Because the LuST Mini scenario has a higher density of vehicles than the Lust scenario, the Kalman filter generated by the CPM contributes to the performance of validation. 
Whereas the proposed method improves the false-positive rate, it also keeps the true-positive rate at the same level as shown in Fig.~\ref{fig:mini_high_TP}.

\begin{table}[tb]
	\begin{center}
		\caption{True- and false-positive rates in the LuST Mini scenario}
		\begin{tabular}{|c|r|r|r|r|} \hline
			& \multicolumn{2}{c|}{Reject Attack [\%]} & \multicolumn{2}{c|}{Reject Valid [\%]}\\
			\cline{2-5}
            & \multicolumn{1}{c|}{Average} & \multicolumn{1}{c|}{Worst} & \multicolumn{1}{c|}{Average} & \multicolumn{1}{c|}{Best} \\ \hline 
            CAM/Front & 91.2 & 80.7 & 45.0 & 40.1 \\ \hline 
            CAM/360 degree & 91.1 & 83.2 & 43.6 & 38.6 \\ \hline 
            CAM\&CPM/Front & 90.4 & 80.7 & 31.0 & 24.4 \\ \hline 
            CAM\&CPM/360 degree & 89.9 & 84.3 & 29.6 & 23.0 \\ \hline 
		\end{tabular}
		\label{table:mini_high}
	\end{center}
\end{table}



\section{Conclusion and future works}\label{sec:conslusion}
Security and privacy are essential in cooperative ITS.
Considering security, CAM is the most fundamental message that conveys information about the surrounding vehicles. 
Therefore, in addition to authentication of nodes with PKI, we need to perform misbehavior detection to deal with attacks against CAM from inside the system.
Nevertheless, misbehavior detection becomes difficult in environments where pseudonym IDs are used for protection against privacy violations caused by tracking.
Our proposed method improves the performance of misbehavior detection by using observation data of other vehicles known as CPM. We experimented in two realistic traffic scenarios and reduced the rate of rejecting valid CAMs (false-positive) by 6 percentage points in the low traffic density scenario and by 15 percentage points in the high traffic density scenario while maintaining the rate of detecting attacks correctly (true positive). 
By improving the usage efficiency of valid CAMs, our method can contribute to traffic safety.

The future studies are as follows. 
The Kalman filter used in this study is suitable for verifying consistency in dynamic system changes; nonetheless, it does not detect quantitative offsets.
In this study, attacks that add quantitative offsets to the original data are out of scope. Considering such attacks, other methods should be used for misbehavior detection.
Regarding the CPM data falsification, we set the limit range of the sensor to minimize its effect in this study. It is necessary to measure this effect and to take countermeasures against the falsification of the CPM data.

\section*{Acknowledgment}
This work was partly supported by JSPS KAKENHI (grant number: 19KK0281 and 21H03423).

\bibliographystyle{unsrt}
\bibliography{shim_paper}

\end{document}